\documentstyle[psfig,emulateapj]{article}
\def\etal{{et~al.}\ }

\def\kms{{\rm\,km/s}}

\def\msun{{\rm\,M_\odot}}

\def\vol#1  {{{#1}{\rm,}\ }}

\def\etal{et al.\ }

\def\clock{\count0=\time \divide\count0 by 60
     \count1=\count0 \multiply\count1 by -60 \advance\count1 by \time
     \number\count0:\ifnum\count1<10{0\number\count1}\else\number\count1\fi}

\begin{document}
\title{Supernovae, Pulsars and Gamma-Ray Bursts: A Unified Picture}
\author{Renyue Cen\altaffilmark{1}}

\altaffiltext{1} {Princeton University Observatory, Princeton University, Princeton, NJ 08544; cen@astro.princeton.edu}

\begin{abstract}

A scenario is proposed that explains both 
the observed high pulsar velocities
and extragalactic gamma-ray bursts (GRBs).
The model involves an ultra-relativistic jet from a supernova (SN),
that produces a GRB and its afterglow,
whose characteristics are similar to an isotropic fireball GRB
perhaps with some differences at late times in the afterglow
once some significant transverse diffusion has occurred.
The time scales and many other properties of 
GRBs and their afterglows in this model are consistent with 
observations.

GRBs in this model have special intrinsic
properties,
that can either falsify or prove this model unambiguously
by observations.
The most direct proof is the detection of a SN
about the same time as the luminous GRB event.
Most GRBs and SNe are expected occur at moderate
redshift ($z\sim 1-3$), if they follow the observed
universal star formation history, as implied in this model.
Searching for GRB/SN associations is a challenge,
because majority of the SNe will be faint.
Some additional, dramatic observable consequences are predicted,
which can also be utilized to test the model.

\end{abstract}


\section{Introduction}

There is by now strong evidence (Metzger \etal 1997; Kulkarni \etal 1998a)
that at least some 
gamma-ray bursts (GRBs) 
are extragalactic in origin. 
While GRBs and 
subsequent ``afterglows" in lower energy bands 
are well explained by the fireball model
(e.g., Rees \& M\'ezs\'aros 1992, RM92 henceforth;
Paczynski \& Rhodes 1993;
Waxman 1997),
the underlying energy source remains a mystery.
In this {\it Letter} I propose a GRB model which 
is directly related to what we know exist and
has direct consequences that can be tested
by current and future observations.

\section{Ultra-relativistic Jet from SN as GRB}

Most SNe are thought to produce strongly
magnetized, rapidly rotating neutron stars,
which later often emit radio pulses (``pulsar"; Manchester \& Taylor 1997).
Gunn \& Ostriker (1970) were the first to recognize 
that Galactic pulsars have a much larger random velocity than
their progenitor massive stars. 
Modern observations of proper motion of pulsars 
indicate even larger velocities with an average of about $450\kms$ 
(Taylor \& Cordes 1993; Lyne \& Lorimer 1994).
Additional supportive evidence comes from observations
of pulsar bow shocks (Cordes, Romani, \& Lundgren 1993;
Predehl \& Kulkarni 1995;
Kaper \etal 1997).
Since not only pulsars in binaries but solitary pulsars
have large velocities, 
it appears necessary
to invoke ``natal kicks" imparted to newborn neutron stars
due to asymmetrical processes during SNe.

Several mechanisms have been suggested
for asymmetrical SN explosions,
including local hydrodynamic instabilities (e.g.,
Herant \& Woosley 1994;
Burrows, Hayes, \& Frywell 1995;
Janka \& Mueller 1996),
global high order gravity modes instabilities 
(Goldreich, Lai, \& Sahrling 1996)
and strong magnetic field induced asymmetries
(Chugai 1984; 
Benesh \& Horowitz 1997;
Lai \& Qian 1998;
Elmfors 1998;
Janka 1998).
While the actual physical cause for the natal kick
is yet unclear,
it is clear that some asymmetries must have also occurred
in the remainder of the SN progenitor besides the neutron star,
in the opposite momentum direction.
It is also clear that the process that produce the asymmetries
should be very swift in the presence of
rapid rotation of proto-neutron stars,
unless the asymmetry aligns with the rotation axis.
Here a concrete picture is conjectured, merely
to illuminate the picture but not to limit possibilities.

It seems natural to hypothesize that such an asymmetry 
aligns with the magnetic axis or the rotation axis (or both)
of the core.
These two axes often
make an angle so as to enable pulsars.
Let us now imagine that the material above
(and perhaps some region below)
the neutrinosphere in a small cone 
(solid angle $\Omega\sim 10^{-2}$sr)
around that special axis is preferentially
first blown out of the deep gravitational potential of 
the star perhaps due to a stronger shockwave in that direction,
which in turn may be caused by a combination of various
physical processes including magnetohydrodynamics, neutrino
dynamics in the presence of strong magnetic field (for example,
parity violation in weak interactions), thermonuclear
and gravitational processes.
Subsequently, when the trapped neutrinos start to escape
from the core,
this narrow cone becomes a preferred exit --- a ``volcano" phenomenon.
A fraction of the neutrinos  
annihilate and convert into an 
electron-positron pair plasma,
which rushes through the cone
carrying with it a small fraction of residual baryonic matter
in the cone.
An ultra-relativistic ``jet" is born.
Whether or not this jet collides with
the material in the cone that was expelled out earlier
has dramatically different consequences, which
will be discussed later.
The subsequent evolution of the jet is similar to 
a spherical fireball model, at least at early stages when
transverse diffusion of the material in the jet can be neglected.

The above depicted mechanism 
for the formation of an ultra-relativistic jet 
is based neither on observational
evidence nor on firm theoretical evidence.
The key element here is to couple a significant
fraction of the total gravitational collapse energy 
of the core (mostly in neutrinos) to a very small amount of baryonic
matter (see below).
But, regardless what the detailed picture of the jet formation 
may turn out to be,
the following analyses of the energetics and rates of such events
as well as the subsequent discussions
of the observational consequences of the resultant GRBs
are, for the most part, independent of such details.

First, let us check the rates of such events.
According to Mao \& Paczy\'nski (1992),
the GRB rate 
is 
$\sim 2\times 10^{-6}$yr$^{-1}$ per $L^*$ galaxy,
and according to Woosley \& Weaver (1986)
the SN rate (for all types of SNe combined) 
is 
$\sim 0.02$yr$^{-1}$ per $L^*$ galaxy.
There are, of course,
differences between different types of SNe.
For instance, in Type Ia SNe core collapse 
is expected not to occur, so GRBs (of the sort described 
here) should not be associated with them.
For simplicity, our analysis below will use
Type II SNe to consider energetics 
and assume that every SN produces a GRB.
It will be simple to rescale energetics, beaming solid angles
and other quantities if only a subset of SNe are responsible for GRBs.
Hence, the beaming solid angle 
is $\Omega\sim 4\pi (2\times 10^{-6}/0.02) \times X\sim 4\pi 10^{-3}~$sr,
merely relating the two rates.
Note that the factor $X$ ($=10$) is inserted somewhat arbitrarily to 
reflect that a large number of faint GRBs may have been missed
by current observations.

Second, the energy budget.
We assume that a fraction, $(\eta+1)\Omega/4\pi$, of the total neutrinos
go through the putative cone, where 
$\eta$ is an enhancement factor ($>1$) since 
the cone is a preferred direction for neutrinos to escape.
We further assume that a fraction $0.003$ (Goodman \etal 1987) of 
the neutrinos turns to a pair plasma.
Then, $\sim 3\times 10^{-3} (\eta+1)(\Omega/4\pi) E_\nu$ will be
carried away by the pair plasma in the cone.
The total neutrino energy in the core is $E_\nu \sim 6\times 10^{53}$erg
(e.g., Wilson \etal 1986 for SN modeling,
consistent with observations of SN1987A, e.g., Bahcall, Spergel, \& Press 1988).
The GRB energy per event is $(\Omega/4\pi) E_{iso}$
(where $E_{iso}=10^{51-53}~$erg; Mao \& Paczy\'nski 1992).
Equating the above two energy terms
indicates that the enhancement factor $\eta$ is of order ten,
a seemingly plausible number.

Third, consider the momentum balance.
For simplicity,
assuming that the remainder of the neutrinos 
escape uniformly in the remaining
solid angle, one has the momentum conservation equation:
${E_\nu\over c} {\Omega\over 4\pi} \eta = m_N v_N$,
where $m_N$ is the neutron star mass,
$v_N$ is the neutron star velocity at birth relative to restframe
of the pre-SN star and $c$ is the speed of light.
We can readily compute $\eta$.
Denoting $m_N=1.4m_{1.4}\msun$, $v_N=450v_{450}$km/s
and $E_\nu=6\times 10^{53}$erg,
one gets 
$\eta=6.3 m_{1.4}v_{450}(\Omega/10^{-3}4\pi)^{-1}$,
consistent with that obtained from energetics requirement.

Fourth, we need to meet the requirement that 
the baryon contaminated jet have
a Lorentz factor $\Gamma\sim 300$,
as needed to yield appropriate 
time scales of GRBs, as observed (see below). 
If the total energy in a GRB is converted
from bulk kinetic expansion energy, the energy requirement gives
$m_B=1.8\times 10^{-5}\left({\Omega\over 4\pi}\right)\left({E_{iso}\over 10^{52}{\rm erg}}\right)\left({\Gamma\over 300}\right)^{-1}\msun$,
the initial baryonic mass of the jet.
This beamed ``fireball" is thus slightly loaded with baryons
(RM92).
It is worth emphasizing that the pre-expel of the baryons in the cone
and the maintenance of its emptiness
are necessary to enable an ultra-relativistic jet,
by noting that the average mass in the cone
is roughly $1.0\msun\times \Omega/4\pi$.
If the small cone does align up or nearly aligns up with the magnetic axis,
in the presence of very strong magnetic field 
magnetic tension may prevent neighboring material
from rapidly diffusing into the emptied cone. 
If the cone also aligns up with the rotation axis,
centrifugal force may also help keep the cone relatively clean.

Last, the relevant time scales.
As in the case of a homogeneous fireball (RM92),
the ultra-relativistic jet cruises until
it has swept up about $1/\Gamma$ 
the original baryon mass $m_B$ (Blandford \& McKee 1976)
at (in comoving frame)
\begin{equation}
t_d=16\hskip -0.1cm\left({E_{iso}\over 10^{52}{\rm erg}}\right)^{1/3}\hskip -0.2cm \left({\Gamma\over 300}\right)^{-2/3}\hskip -0.2cm \left({n\over 1 {\rm cm}^{-3}}\right)^{-1/3}\hskip -0.1cm{\rm days},
\end{equation}
\noindent
corresponding to a radius of 
\begin{equation}
r_d = 2.8\times 10^{16}\hskip -0.1cm\left({E_{iso}\over 10^{52}{\rm erg}}\right)^{1/3}\hskip -0.2cm \left({\Gamma\over 300}\right)^{-2/3}\hskip -0.2cm \left({n\over 1 {\rm cm}^{-3}}\right)^{-1/3}\hskip -0.1cm{\rm cm},
\end{equation}
\noindent
where $n$ is the number density of the circumburst medium.
Thereafter, $\Gamma$ decreases rapidly 
as $r^{-3}$ (radiative expansion) or $r^{-3/2}$ (adiabatic expansion).
Hence, most of the bulk energy in the jet is thermalized
and radiated away in $\gamma-$ray
at $\sim r_d$ (in local comoving frame)
within a distance of $\sim r_d$ (in local comoving frame) 
in a time duration in observer's frame (RM92):
\begin{equation}
\Delta t_{GRB} \sim {r_d\over 2\Gamma^2 c} = 4.8\left({E_{iso}\over 10^{52}{\rm erg}}\right)^{1/3}\hskip -0.2cm \left({\Gamma\over 300}\right)^{-8/3}\hskip -0.2cm \left({n\over 1 {\rm cm}^{-3}}\right)^{-1/3}
\end{equation}
\noindent
seconds, in accord with the durations of observed GRBs of $\sim 1-100~$sec 
(Norris \etal 1995).
For a GRB at redshift $z$, $\Delta t_{GRB}$ will be boosted by
a factor of $(1+z)$.


\section{Testable Features of Proposed GRBs}

This model has some unique observable properties.
The most direct test is that
a SN sets off (i.e., the moment of the onset of SN neutrino burst)
about the same time as the GRB.
There are several possible outcomes.

1) In the event that the jet is unblocked by 
   the pre-expelled material in the cone and 
   is in the line of sight, a ``normal" energetic GRB is seen with
   energy release in $\gamma$-ray of the isotropic equivalent
   of $E_{iso}\sim 10^{51-53}$erg.
   The fast rotation of the neutron star makes possible for 
   the pre-expelled meterial not to block the jet,
   if the cone is not aligned with the rotation axis.
   GRB970508 
   and GRB971214 
   belong to this category.
   The companion SN, if observed,
   may be somewhat ``atypical" 
   because of the incoming SN ejecta
   comes from a somewhat special direction in the vicinity
   of the jet.
   The spatial location of the SN should coincide with that of the GRB.

2) If the jet is partially blocked by the pre-expelled material
    when it is aligned with or close to the rotation axis,
    several things could happen.
    Some of the pre-expelled material will be accelerated to a velocity 
    much higher than usual SN ejecta velocity.
    But due to the smallness of the baryonic mass in the jet
    compared to the mass in the ejecta,
    the ejecta is still accelerated only to a sub-relativistic speed.
    The collision takes place near the surface of the star
    and the portion of the jet that involves in the collision
    will be immediately decelerated to a sub-relativistic speed,
    producing super-MeV $\gamma$-rays at the same time.
    However, no significant amount of 
    $\gamma$-rays can escape due to high pair opacity
    in the small involved region (Goodman 1986).
    Therefore, almost all of the energy of the collided portion of the jet 
    turns into expansion energy of the massive ``ejecta".
    We can, however, still see the remaining portion of the jet that
    is unblocked, with its energy scaled
    downed by a factor proportional to the fraction 
    of the solid angle occupied by the unblocked part of the jet,
    {\it if it happens to be in the line of sight}.
    GRB980425 may be such an event where only a small fraction
    ($10^{-3}-10^{-2}$) of the jet escapes.
    The associated SN, SN1998bw, should be unusually bright,
    just as observed (Kulkarni \etal 1998b; K98 henceforth),
    which led K98
    to conclude that SN1998bw is expanding at a relativistic speed.
    We argue that the radio observations of SN1998bw might be 
    explained if the radio signal is a blend of a SN
    and the afterglow from 
    a weak GRB, 
    which is still very bright in SN standard.
    Intriguingly, the ``new" radio component of SN1998bw as
    observed by K98 (their Figure 2)
    may be due to the radio afterglow of GRB980425.
    This removes the enormously demanding energy requirement 
    that SN1998bw is expanding at a relativistic speed
    ($E_{ejecta} = \Gamma_{ejecta} m_{ejecta} c^2 \sim 10^{54}$erg).
    The high $\Gamma$ of the GRB jet 
    is also consistent with the radio scintillation observations
    (K98), which show small temporal fluctuations thus
    a large apparent size.
   The spatial location of the SN should also coincide with that of the GRB 
   in this case.

3) When only a small portion near the edge of the jet enters
   the line of sight, a weak GRB would result, accompanied by 
   a SN. The properties of such a GRB/SN event should be 
   similar to 2) and could also
   account for the observed GRB980425/SN1998bw event.
   The SN should be coincidental with the GRB spatially.

4) When the GRB beam is completely off line of sight,
one should still expect some consequences.
We should still see a Compton scattered GRB event with a comparable
duration but with a much lower luminosity and significant polarization, 
whose energy spectrum should be softer than direct GRBs (cases 1,2,3).
However, an unique feature occurs here.
The GRB event will be delayed by 
weeks compared to the SN event (eq. 1) and
the GRB should be spatially displaced in the sky
by light-weeks (eq. 2).
The subsequent afterglows will be further delayed and
spatially displaced from the SN.
Combining the 
time delay and the angular separation between the SN and GRB
(probably only for nearby ones in practice)
should allow a determination of
the angle of the jet relative the line of sight,
which, when combined the observed (scattered) spectrum,
should further allow a recovery of the original spectrum.
The vast majority of GRB/SN events belong to this class
and it is urgent to look for them.
It is intriguing to note that the ``mystery spot" or bright ``companion"
to Supernova 1987A
may well be such an event.
Interestingly, the time delay of the IR ``echo" in SN1987A
of $\le 30~$days is remarkably close to the {\it projected}
separation between SN1987A 
and its ``companion" of $19~$light-days
(Felten, Dwek, \& Viegas-Aldrovandi 1989).
This is completely consistent with an ultra-relativistic jet which 
leaves the base (neutron star) about the same time as the
SN explodes and whose back is seen obliquely at the indicated time.

Some other observable features of this model may also be predicted.
Here we list several of them.

First, assuming
the total energy output from each jet is roughly constant,
longer GRBs due to blastwaves propagating into lower 
density interstellar medium should be fainter,
in good agreement with observations (Kouveliotou \etal 1993).
A broad anti-correlation between peak luminosity and
duration 
should result,
with a significant scatter due to the expected
broad redshift distribution.
There may be some correlation between peak energy 
and luminosity (Mallozzi \etal 1995)
or duration (Fenimore \etal 1995):
longer and fainter bursts 
should be softer spectrally due to a combined effect
of lower $n$, perhaps lower magnetic field and cooler electrons.
It is also expected that
longer duration GRBs outnumber
shorter duration GRBs,
since the filling factor
of lower density regions in the interstellar medium
is normally larger than 
that of high density regions,
in agreement with observations (Hakkila \etal 1996).
Also, it is expected that there  should be 
a variety of temporal profiles in both GRBs 
and their afterglows, due to 
complex circumstellar and interstellar medium distributions.
Finally, if more than one types of SNe are
responsible for the observed GRBs,
the above correlations may still hold but with 
substantially larger scatters.

Second, 
GRB repeaters might be possible if very empty voids
exist in the interstellar medium.
The pause between two successive bursts
would correspond to the time for the jet to travel through an
empty void between two dense regions (along the line of sight),
during which the jet suffers no slowdown thus no significant
loss of energy.
The time interval between successive bursts
provides a measure of void size.
The October 27-29, 1996 
four consecutive bursts in the same sky location (Connaughton \etal 1997)
are interesting in this respect.
They consist of two pairs,
18 minutes and 11.2 minutes apart, respectively,
separated by 1.8 days.
In the framework of the model proposed here,
the distances that the jet travels
and times elaped in the jet comoving frame (using $\Gamma=300$)
in the three respective time intervals
are ($4\times 10^{18}$cm, $6\times 10^{18}$cm, $8\times 10^{20}$cm)
and ($4$yrs, $6$yrs, $900$yrs),
compared to the typical size of a star formation region 
(a giant molecular cloud) of $\sim 50$pc.
Thus it seems that this model is capable of 
producing the two pairs of bursts with smaller time intervals as 
repeaters from a single jet but difficult to accommodate
repeaters separated by $1.8~$days. 
For the same reason,
a diversity of GRB afterglows may be expected.
Many GRBs may lack afterglows (not just in optical)
(e.g., Feroci \etal 1998 for GRB970111).
This may result, for instance, in a situation
where the jet does not encounter a significant amount of matter
for a prolonged period of time subsequent to the GRB.
Thereafter, even if the jet encounters some significant amount of matter,
a weaker magnetic field and/or lower density (among other factors)
may render the afterglows faint and/or at much lower energy frequencies.

Third, since massive stars 
do not move significantly from their birth places before exploding,
the GRBs 
are predicted to occur {\it only in star-forming regions.}
The latest observations might have provided some evidence for this 
as summarized by Paczy\'nski (1998 and references therein)
and newer evidence from GRB971214 optical afterglow reddening
(Halpern \etal 1998).
There might be exceptions 
when one massive star in a binary
explodes first and kicks the neutron star-massive star binary
out of a star-forming region.
But the maximum
distance that the binary travels away from the star-formation region
before the second massive star explodes as a SN
is roughly $10^6$yrs$\times 450v_{450}$km/s$\sim 0.46v_{450}$kpc.
So, it is unlikely to see any GRB in halos of large, luminous galaxies
(sighting GRBs in outskirts of small galaxies
is not impossible).
This feature and the two that follow
are shared by some other models with GRB hosts 
in star-forming regions (e.g., the hypernova model
of Paczy\'nski 1998).

Fourth, since GRBs occur in star-forming regions,
many GRBs may exhibit weak or lack optical afterglows
due to large dust extinction 
in star-forming regions (e.g., van den Bergh 1992;
Hanson, Howarth, \& Conti 1997;
Whitney, Kenyon, \& Gomez 1997).
While self-cleaning due to the UV flux from the massive stars
themselves would remove most of dust obscuration in 
the immediate vicinity of the massive stars (hence GRBs),
most of the dust obscuration probably lies 
along the line of sight.
A search for
afterglows in infared might turn out to be fruitful.

Fifth, the redshift 
distribution of GRBs should follow
the star-formation history (Madau 1997).
Eventually, the distribution of GRBs in the sky
should correlate strongly with galaxies, {\it once observations
will be sensitive enough to encompass most galaxies, i.e.,
reach beyond the redshift peak of galaxies ($z\sim 1$)}.
There should virtually be no correlation between luminous GRBs 
any local extragalactic or galactic sources.
The probability of seeing association of SNe in Local Group 
galaxies with bright GRBs (as described here)
is small, $<10^{-4} (t_{obs}/{\rm yrs})$, where
$t_{obs}$ is the total amount of time used to monitor GRBs.
However, faint, scattered GRB events (case 4 above) 
should be strongly correlated with the Galaxy and nearby galaxies and
it is urgent to search for these.

\section{Conclusions}

A unified model that explains both high pulsar velocities
and cosmological gamma-ray bursts is proposed.
The model involves an ultra-relativistic jet
from a SN. 
The model has several directly
testable observational consequences,
as detailed in the previous section.
In particular, detection of an associated SN that sets off 
(onset of SN neutrino burst) 
about the same time as the GRB 
in the event that both the GRB and SN are seen 
presents a direct test of the model.
It is noted that the evolution of the jet 
may be dramatically modified in the presence 
of strong magnetic field in near equipartition
due to magnetic confinement
compared to otherwise (e.g., Rhoads 1997).
Off line of sight GRBs, which constitute a vast majority,
also have interesting and dramatic 
observable consequences which could provide additional
tests of the model.

The theoretical basis for this scenario is 
uncertain and the model is at best suggestive.
Nevertheless,
it is based on
one set of known energetic events (supernovae) and also accounts
for the motion of another set of known objects (pulsars).

After this paper was submitted,
a paper by Wang \& Wheeler (1998)
appeared at the LANL preprint site,
which presents a similar scenario for GRBs.
The major difference between their scenario and the scenario
presented here
is that they associate the low brightness GRBs such as GRB980425
with the isotropic component of the SN ejecta and high brightness
GRBs such as GRB971214 with the jet,
whereas we claim that both types of GRBs 
are due to the jet, occuring or being viewed under different conditions.

\acknowledgments
The work is supported in part
by grants AST9318185 and ASC9740300.
Discussions with Jeremy Goodman and
Russell Kulsrud are gratefully acknowledged.


\end{document}